# The assertive profile of the Bulgarian students in computer science and computer engineering

Ivelina Peneva[a], Krasimir Yordzhev[b]

[a] *South-West University, Faculty of Philosophy, 66 Ivan Mihailov Str, 2700, Blagoevgrad, Bulgaria*
E-mail address: *ivelina_peneva@swu.bg*

[b] *South-West University, Faculty of Mathematics and Natural Sciences, 66 Ivan Mihailov Str, 2700, Blagoevgrad, Bulgaria*
E-mail address: *yordzhev@swu.bg*

**Abstract:** Different points of view on the nature and content of the assertiveness are followed in this paper. The main purpose is to study the assertive profile of Bulgarian students in computer science and computer engineering by analyzing the components of assertiveness. Research was performed using testing methods. It was found that the level of expressivity of this personal quality among subjects were above-average level.

## 1. Introduction

Particular attention was paid to the role of assertiveness in the modern educational system. Many authors accepted the assertive behavior as the most constructive method of communication in the teaching process and its application in the teaching practice not only increased the effectiveness of training activities, but it also had a positive educational impact on young people, facilitating the development of valuable personal qualities in them [1, 2].

Special attention was paid to the formation of assertive skills in upper grade and college students, as it was believed that this age was conducive to build a vision, perseverance, self-reliance and initiative, qualities that were objective prerequisite for development of assertiveness. The development of these characteristics in learners improved the performance of communicative interaction with teachers and peers and influenced the effectiveness of learning by reducing stress in school, increasing the success rate of trainees, and expanding their reflective learning [3, 4].

The purpose of this paper is the study of assertive profile of Bulgarian students in computer science and computer engineering.

Although by its nature psychic phenomena (including assertiveness) are determined by random events and have a probabilistic nature [5], it is expected to indicate some general trends among the group of subjects tested, given their professional orientation and motivation for future career.

## 2. Essence of the assertiveness as a psychic phenomenon

Concept of assertiveness is a relatively recent entrant in the psychological literature and practice. It derives from the English word *assert*, that means declare, affirm, prove. There are differences on the interpretation of the essence of this concept. Some authors use it as a synonym for self-sufficiency and confidence [6]. E. Solter accepts assertiveness of personal feature that can be defined as autonomy and independence from external influences, ability to self-regulation of behavior [7]. Assertiveness, according T. Paul presumes free behavior consistent with own motives and beliefs and regardless of somebody else's opinion. J. Wolpe believes that the main manifestation of assertiveness is the sense of self-esteem and ability for active orientation in life [8]. These authors understand assertiveness as a confidence, a willingness to take independent decisions regarding their own lives and taking responsibility for them, i.e., they emphasize the attitude of the individual to it.

Other psychologists add to that the respect and acceptance of others. R. Frichie [9] believes that assertive is a person who is responsible for his own behavior, who demonstrates respects for others and seeks compromises. Sue Bishop [10] defines assertiveness as a form of behavior based on self-responsibility and awareness of the rights of others, thereby demonstrating self-respect and respect for others. That very last point, according to St. Stein [11] distinguishes the assertiveness from aggression which also aims to achieve self-interest, but not taking into consideration, and violating the rights of others. This position sees the assertiveness as the optimal and most constructive way of interpersonal interaction, founded on the principles of humanism, thereby denying manipulation, cruelty and aggression in relation to other people and avoiding injury and lesion of the rights of the interacting parties. Some authors focus precisely on these personal rights. Patricia Jakubowski defines assertiveness as an act of protecting their own human rights without to infringe upon of the fundamental rights of others [12]. This is a style of response that takes into account the limits between the rights of an individual and the rights of others and acts so as to preserve these limits stable. M. Smith defines assertive person as a person aware of his rights, exercising and defending them when necessary. At the same time the assertive person recognizes the rights of others and seeks in their relationships not violate them.

This author is known for the so - called**,** Code of assertiveness - list of 10 personal rights which include "the right to judge your own behavior, thoughts and emotions," "the right to be independent of the opinions of others," "the right to make mistakes" and others [13]. Similar lists of assertive rights are prepared by another authors dealing with the issues of assertiveness [10, 14, 15, 16]. Basically they pay attention to the freedom of a person to express his feelings and opinions without uneasiness and depending on the approval of others, to have different necessities and needs than other people, to be critical of others in a constructive way, as well as to face up to criticism towards him, to make his own decisions and deal with the consequences, to change his mind and be independent in his intentions and actions**.**

Summarizing the concepts of individual authors assertiveness can be defined as a complex multiple personable construct, the main components of which are the presence of self-confidence, self-respect and respect for others, a person's ability to actively defend its interests openly declares its objectives, intentions and feelings, not impaired in the interests of the surrounding. O. Fedosenko [6] structures qualities of assertiveness, identifying four components - cognitive, affective, motivational and behavioral.

**The cognitive component** consists of the openness of acquiring knowledge about you, others and the situation which leads to adequate orientation and assessment of the circumstances of life and one's own behavior.

**Affective component** includes self-confidence, self-respect and respect for others and accepting their position and rights. Also assertive person should be well aware of his emotions and express they openly, whereby not hurt feelings of others.

**Motivational component** is oriented towards solving the tasks corresponding to the actual level of abilities and pursuit of achieving the best results.

**Behavioral component** manifests itself in independence, high level of self-control and confidence in solving life and professional tasks.

### 3. Research program

The purpose of this paper is the study of assertive profile by analyzing the components of assertiveness in students in computer science and computer engineering at the Faculty of Mathematics and Natural Sciences of SWU " Neofit Rilski " - Blagoevgrad.

The survey was conducted in late May 2012 among 84 students aged 19-24 from them girls are 33, boys are 51. They have no problems with the acquiring of the material; the majority has good success in the first term. Among them there are no persons with antisocial behavior or behavioral abnormalities. No suffering from chronic illnesses, including mental health.

The study used methods:
1. Scale „Your assertiveness profile", developed by Judith Tindall [16];
2. Test measuring the emotional intelligence, developed by N. Schutte [17];
3. Rosenberg self-esteem scale [18];
4. Social desirability scale, created by D. P. Crowne and D. Marlowe [19];
5. W. Fey's scales Acceptance of others [20];
6. Test measuring the motivation for achievement, created by I. Paspalanov and I. Shtetinski [21];
7. Rotter's Locus of Control Scale [22];
8. Self-Monitoring Scale (SMS), created by Mark Snyder [23].

### 4. Results from the study

Table 1 shows the results of the applied test for the study of assertiveness in students (method 1).

Table 1. Results of the personal scale for researching of assertiveness.

| level of markedness (in % subjects tested) | girls | boys | total subjects tested |
|---|---|---|---|
| low | 0 | 0 | 0 |
| medium | 51,2 | 56,0 | 53,5 |
| high | 48,8 | 44,0 | 46,5 |

Average value of the indicator of assertiveness was 48.6 (60 %). This is a relatively high result and corresponds to the behavior in which is common assertive behavior. As a positive fact we can note that none of the tested individuals have lower levels of assertiveness. Most of them have a medium level of markedness of this quality (53,5 %). Girls have higher results than those of boys, but the differences were not statistically significant (Table 6).

For establishing cognitive capabilities of the individual is usually used classical intelligence test. The following was not considered as necessary to be applied in this study, as there is sufficient research in this area, including on students. It is interesting a little studied component of intelligence - emotional intelligence. Although it concerns the affective side of the human psyche it is based on cognitive mechanisms to ensure the processing of emotional aspects of personality. In this sense, the study of emotional intelligence will provide important information about the ability of the surveyed students to recognize and understand their own emotions and emotional behavior and those of others.

Table 2 shows the results of the applied test for the study of emotional intelligence (method 2).

Table 2. Distribution of the surveyed students (in %) who received low, medium and high level of emotional intelligence test.

| level of markedness | Total score | factor I – „Sharing of emotions and empathy" | factor II – „Motivation for the overcoming of difficulties and optimism" | factor III – „Recognizing of other people's non-verbal expression of emotions" | factor IV – „Recognition of own emotions and self-control" |
|---|---|---|---|---|---|
| low | 29,8 | 45,2 | 31,0 | 4,8 | 25,0 |
| medium | 64,3 | 50,0 | 63,1 | 91,4 | 67,9 |
| high | 6,0 | 4,8 | 6,0 | 4,8 | 7,1 |

It is estimated the prevailing medium level of markedness as the overall test as well as the individual factors. There is a relatively high percentage (up to 45,2% depending on factors) of students with low markedness of the qualities that indicates poor ability to recognize emotions, their display, sharing and controlling.

To study the affective (emotional) component of assertiveness are applied methods 3, 4 and 5 for the research of self-esteem, social desirability and acceptance of others (Table 3).

Table 3. Results from a study of the emotional component of assertiveness (in % subjects tested).

| level of markedness | Self-esteem | Social desirability (need for approval) | Acceptance of others |
|---|---|---|---|
| low | 6,0 | 14,3 | 4,8 |
| medium | 70,2 | 70,2 | 79,8 |
| high | 23,8 | 15,5 | 15,5 |

It is estimated the prevailing average level of self-esteem of students, only 6 % have low levels of this indicator. Good self-esteem and self-confidence are associated with lower levels of need for approval and conformity of the students, future computer professionals. Over 85% of those surveyed could not comply with the commonly accepted opinion if it contradicts their own views. At the same time the students accept and respect the opinion of others, they tolerate and take into consideration their rights, which are indicated by the high results in the test " Acceptance of others".

**Motivational component** of assertiveness is examined through the test of motivation for achievement (method 6), the data of which are shown in Table 4.

94% of the students are medium and high motivated to achieve distant perspectives and they are consistent and persistent in achieving their goals. Over 90% are the students who are oriented towards high standard of performance and who have business interests in their activity (cluster D and E).

As an indicator for study of the motivational component of assertiveness can serve also factor II from the test of emotional intelligence, which concerns the motivation for the overcoming of difficulties (Table 2). According to the results nearly 70% of the students are focused and motivated in their activity and they are optimistic about their final results and achievements.

Table 4. Distribution of the surveyed students (in %) who received low, medium and high level of test motivation of achievement.

| level of markedness | Total score | cluster D "Business Orientation" | cluster E "Orientation to a high standard of performance" |
|---|---|---|---|
| low | 6,0 | 4,8 | 6,0 |
| medium | 72,6 | 73,8 | 70,2 |
| high | 21,4 | 21,4 | 23,8 |

**Behavioral component** of assertiveness is examined through the methods 7 and 8, respectively, measuring the level of self-regulation and self-control in communication. According to the results 51,2% of the students have internal locus of control, i.e. they are able to adequately control their behavior, thoughts, and external manifestations of their emotions. Only 7.7% had low levels of communicative control (Self-Monitoring Scale). The others are able to react flexibly to changes in a given situation and to adapt quickly to different social circles (Table 5).

Table 5. Results from a study of the behavioral component of assertiveness (in% subjects tested).

| level of markedness | Locus of control | level of markedness | Self-Monitoring |
|---|---|---|---|
| external | 48,8 | low | 6,0 |
| internal | 51,2 | medium and high | 94,0 |

The degree of independence from external evaluations among the students can be estimated by the levels of need for approval by the Scale of social desirability (Table 3). Only 15,5 % had high results on this indicator, the majority of the surveyed students do not feel dependent on the opinions of others and they are able to define and manage their behavior according to their own views. To establish statistically significant differences between the sexes in the studied parameters was applied Student's t-test. The results are shown in Table 6.

Table 6. Differential gender differences in individual parameters examined.

| indicator | sex | M | t | df | p |
|---|---|---|---|---|---|
| Assertiveness | male | 55,14 | 0,133 | 82 | 0,447 |
|  | female | 54,88 |  |  |  |
| Emotional intelligence | male | 113,15 | 0,378 | 82 | 0,353 |
|  | female | 115,42 |  |  |  |
| Self-esteem | male | 27,30 | 2,959 | 82 | 0,002 |
|  | female | 31,0 |  |  |  |
| Social desirability | male | 14,35 | 1,235 | 82 | 0,112 |
|  | female | 15,52 |  |  |  |
| Acceptance of others | male | 21,50 | 1,016 | 82 | 0,158 |
|  | female | 22,57 |  |  |  |
| Motivation for Achievement | male | 10,35 | 1,416 | 82 | 0,082 |
|  | female | 11,57 |  |  |  |
| Locus of Control | male | 17,48 | -1,444 | 82 | 0,078 |
|  | female | 16,14 5,45 |  |  |  |
| Self-Monitoring | male | 4,10 | -1,494 |  | 0,072 |
|  | female | 3,31 |  |  |  |

There are no significant differential gender differences. There are only estimated lower levels of the test for self-respect among boys compared to the girls. This is not drastically downwards and should not be interpreted as a trend towards reduced confidence in boys and an indicator of unbalanced self- esteem. According to the percentage distribution of the test results for self- respect predominant (over 90%) among youth are the medium and high levels of markedness of this indicator.

## 5. Summary

This study allowed revealing some specific characteristics of the assertive profile of students who are going to be future computer professionals. There were estimated quite positive trends in their personal development. Above all, the general level of assertiveness is above average, more over there aren't subjects tested with low performance on this criterion.

As regards of the **cognitive component** of assertiveness are estimated prevailing average values of the studied emotional intelligence. Relatively high percentage of low level of that indicator should not be considered as a negative trend. Rather, it is a normal consequence of the prevailing rationality in people dealing with science compared to emotions.

As regards of the **emotional component** are estimated exceptionally high results. Over 95% of the tested subjects have medium and high levels of self-respect and acceptance of differences in others. The students with conformal characteristics are few nearly 90% are independent of the opinions of others in their views and behavior.

The indicators in the **motivational component** of assertiveness are high as well. The majority of the students are focused and persistent in achieving their goals, they are aspiring to significant achievements and reach them through high standards in their actions.

In the study of the **behavioral component** of assertiveness good results were estimated as well. Most of the students are confident and independent on external evaluations and influences and they are able to control their thoughts and behavior.

## 6. Conclusion

The summarized results of the complex study shows that the students in computer science and computer engineering are highly assertive, they have good emotional and communicative culture, they are highly motivated and able to reach their goals through skillful control of their activity and emotions. These results are an indicator of social and personal maturity of the tested subjects and ensure their future professional achievements and personal success.

The results of the study provided an opportunity to explore the assertive profile of students who are going to be future computer professionals. We consider that this has not only theoretically, but above all practical significance, since it furthers more specific estimation of the means for focused influence to correct negative tendencies and to develop the positive aspects in the personal structure of the students.